\documentclass[fleqn,usenatbib]{mnras}

\usepackage{newtxtext,newtxmath}

\usepackage[T1]{fontenc}

\DeclareRobustCommand{\VAN}[3]{#2}
\let\VANthebibliography\thebibliography
\def\thebibliography{\DeclareRobustCommand{\VAN}[3]{##3}\VANthebibliography}

\usepackage{graphicx}	
\usepackage{amsmath}	

\title[Youngest Pulsar Wind Nebula in SN 1986J]
{
	Testing a stochastic acceleration model of pulsar wind nebulae: \\
	Early evolution of a wind nebula associated with SN 1986J
}

\author[S. J. Tanaka \& K. Kashiyama]{
Shuta J. Tanaka,$^{1,2}$\thanks{E-mail: sjtanaka@phys.aoyama.ac.jp (SJT)}
and
Kazumi Kashiyama$^{3,4}$
\\
$^{1}$Department of Physical Sciences, Aoyama Gakuin University, 5-10-1 Fuchinobe, Sagamihara, Kanagawa 252-5258, Japan\\
$^{2}$Graduate School of Engineering, Osaka University, 2-1 Yamadaoka, Suita, Osaka 565-0871, Japan\\
$^{3}$Astronomical Institute, Graduate School of Science, Tohoku University, Aoba, Sendai, Miyagi 980-8578, Japan\\
$^{4}$Kavli Institute for the Physics and Mathematics of the Universe (Kavli IPMU,WPI), The University of Tokyo, Kashiwa, Chiba 277-8582, Japan
}

\date{Accepted XXX. Received YYY; in original form ZZZ}

\pubyear{2023}

\begin{document}
\label{firstpage}
\pagerange{\pageref{firstpage}--\pageref{lastpage}}
\maketitle

\begin{abstract}
    Over three thousand pulsars have been discovered, but none have been confirmed to be younger than a few hundred years.
    Observing a pulsar after a supernova explosion will help us understand the properties of newborn ones, including their capability to produce gamma-ray bursts and fast radio bursts.
    Here, the possible youngest pulsar wind nebula (PWN) at the center of the SN 1986J remnant is studied.
    We demonstrate that the 5 GHz flux of `PWN 1986J', increasing with time, is consistent with a stochastic acceleration model of PWNe developed to explain the flat radio spectrum of the Crab Nebula.
    We obtain an acceleration time-scale of electrons/positrons and a decay time-scale of the turbulence responsible for the stochastic acceleration as about 10 and 70 years, respectively.
    Our findings suggest that efficient stochastic acceleration and rising radio/submm light curves are characteristic signatures of the youngest PWNe.
    Follow-up {\it ALMA} observations of decades-old supernovae within a few tens of Mpc, including SN 1986J, are encouraged to reveal the origin of the flat radio spectrum of PWNe.
\end{abstract}




\section{Introduction}\label{sec:intro}

Neutron stars are thought to be left behind supernova (SN) explosions \citep[][]{Baade&Zwicky34} and this idea is strongly supported by the detection of the neutrinos associated with SN 1987A \citep[][]{Hirata+87,Bionta+87}.
On the other hand, the existence of neutron stars themselves is confirmed by the discovery of radio pulsars \citep[][]{Hewish+68} and the remarkable example of the Crab pulsar and its nebula, which is the supernova remnant (SNR) of SN 1054, makes the association between SN explosions and neutron stars no doubt \citep[e.g.,][]{Hester08}.
Nevertheless, we have never observed a pulsar subsequent to an SN explosion.
The most studied case must be the remnant of SN 1987A, but the detection of the emission from the compact object at the center is under debate \citep[][]{Alp+18, Alp+21, Greco+21}.

The fundamental question that is missing in the above picture is when the pulsar mechanism starts to work \citep[][]{Goldreich&Julian69, Ruderman&Sutherland75} and also how large is the spin-down power and energy of a newborn pulsar.
These very questions are shared in the context of the super-energetic astrophysical phenomena, such as long-lasting activities of gamma-ray bursts \citep{Zhang&Meszaros01}, luminous supernovae \citep[e.g.,][]{Chevalier&Fransson92, Metzger+14, Murase+15, Kashiyama+16, Hatsukade+21}, and also fast radio bursts \citep[e.g.,][]{Kashiyama&Murase17, Kisaka+17}.
In all the above cases, the newborn neutron star has to work as a pulsar immediately after its formation in order to energize the phenomena, so that they must be rapidly rotating ($\sim$ msec) and/or highly magnetized ($\gtrsim 10^{14}$ G) pulsars \citep[e.g.,][]{Dall'Osso&Stella21}.
Note that such energetic pulsars are far away from the known pulsars ever observed on the $P - \dot{P}$ diagram, and especially from the Crab pulsar \citep[e.g.,][]{Faucher-Giguere&Kaspi06}.
The Crab pulsar is the only pulsar whose birth spin period and magnetic field can be determined from observations under the constant braking index assumption.

SN 1986J is classified as an SN IIn \citep[][]{Rupen+87} and is known as a candidate for hosting a 30 years-old pulsar.
SN 1986J was first discovered in radio about three years after its explosion and is located in the nearby edge-on galaxy NGC 891 \citep{vanGorkom+86, Chevalier87, Rupen+87, Weiler+90}.
\citet{Bietenholz+04} discovered a spatially and spectrally distinct component, the central component, in addition to the expanding (moving) shell component with Very-Long-Baseline Interferometry (VLBI) \citep[][]{Bietenholz+02}.
\citet{Bietenholz&Bartel17a} showed that the 5 GHz light curve of the central component increases with time, while that of the shell component decreases.

In this paper, we interpret the central component of SN 1986J as the pulsar wind nebula (PWN) powered by the remnant pulsar of SN 1986J (hereafter, PWN 1986J and PSR 1986J, respectively).
The remnant pulsar and its nebula are assumed to be Crab-like and the observed light curve is studied based on the stochastic acceleration model developed by \citet{Tanaka&Asano17}.
The stochastic acceleration model considers that the turbulence driven by the interaction between the pulsar wind and the SN ejecta accelerates the externally injected non-relativistic particles and forms a population of the radio-emitting particles distinct from that of the X-ray-emitting particles directly injected from the central pulsar.
The stochastic acceleration model is a promising solution in order to resolve the origin of both the observed flat radio spectrum much harder than the X-ray one and the observed electron/positron pair amount inside PWNe much larger than theoretical predictions \citep[see also][]{Tanaka&Takahara13b, Timokhin&Harding15, Xu+19, Lyutikov+19}, while the conventional broken power-law injection model does not account for the amount of the radio-emitting particles \citep[e.g.,][]{Bucciantini+11, Tanaka&Takahara11, Tanaka&Takahara13a}.
The presence of the strong turbulence would also provide a solution to the classical problem of PWNe known as the $\sigma$-problem \citep[][]{Kennel&Coroniti84, Porth+14, Zrake&Arons17, Tanaka+18}.

Here, we will show that a characteristic behavior of the stochastic acceleration model can also reproduce the radio flux increase observed from PWN 1986J.
The stochastic acceleration time-scale of the relativistic turbulence is reflected in the rising radio light curve in this model.
This is a different picture from the conventional broken power-law model, where the observed radio flux increase corresponds the transition phase from the optically thick to the thin regime (see section \ref{sec:Dis&Cons}).
In section \ref{sec:Model}, we describe our model which is slightly updated from \citet{Tanaka&Asano17}.
In section \ref{sec:Application}, we present the results of the application to the case of SN 1986J.
Section \ref{sec:Dis&Cons} is devoted to the discussion and the summary of the present paper.

\section{Model}\label{sec:Model}

The present model is a one-zone spectral evolution model of a spherical PWN expanding within its parent SNR.
The dynamics of the PWN expansion is briefly described in section \ref{sec:Expansion}.
For the spectral evolution, we adopt the stochastic acceleration model developed by \citet{Tanaka&Asano17}.
This model gives a plausible explanation for a `flat radio spectrum', which is a characteristic of PWNe, and we will see that the same model can also explain the radio behavior of SN 1986J.
In section \ref{sec:Acceleration}, we summarize the source of the large amount of radio-emitting particles and also the way how the hard energy spectrum of them is formed.
In section \ref{sec:FittingProcedure}, we summarize the parameters of this study and describe the fitting procedure.

\subsection{Expansion of SNR and PWN}\label{sec:Expansion}

The evolution of the forward and reverse shocks of a SNR (energy $E_{\rm SN}$ and mass $M_{\rm ej}$) surrounded by a uniform interstellar medium of a mass density $\rho_{\rm ISM}$ is given by \citet{Truelove&McKee99}.
The expansion of a PWN within a parent SNR is calculated by the thin-shell approximation and then the radius of a uniform spherical PWN is equal to the radius of the shell $R_{\rm sh}$ \citep[c.f.,][]{Gelfand+09, Bandiera+20}.
The PWN sweeps up the tail part of the expanding SN ejecta material by forming a massive shell with a velocity $v_{\rm sh}(t) \equiv \dot{R}_{\rm sh}(t)$ and of a mass $M_{\rm sh}(t)$, which are given by
\begin{eqnarray}
	\frac{d}{d t} M_{\rm sh}(t)
	&\equiv &
	\dot{M}_{\rm sh}(t)
	=
	4 \pi R_{\rm sh}(t)^2 \rho_{\rm ej}(R_{\rm sh}(t),t) \nonumber \\
	& \times & \left[ v_{\rm sh}(t) - v_{\rm ej}(R_{\rm sh}(t), t) \right], \label{eq:MdotShell} \\
	\frac{d}{d t} M_{\rm sh}(t) v_{\rm sh}(t) & = & 4 \pi R_{\rm sh}^2(t) \left[ P_{\rm PWN}(t) - P_{\rm ej}(R_{\rm sh}(t), t) \right] \nonumber \\
	& + & v_{\rm ej}(R_{\rm sh}(t), t) \dot{M}_{\rm sh}(t), \label{eq:ShellEOM}
\end{eqnarray}
where $P_{\rm PWN}(t)$ and $P_{\rm ej}(r,t)$ are the pressure of the uniform PWN and the SN ejecta described as follows.
The hydrodynamic properties of the SN ejecta in early phase are taken from \citet{Blondin+01}, i.e.,  $v_{\rm ej}(r, t) = r / t$, $P_{\rm ej}(r, t) = 0$, and
\begin{eqnarray}\label{eq:EjectaDensity}
	\rho_{\rm ej}(r,t) & = &
	\frac{5(\omega - 5)}{2 \pi \omega \pi} E_{\rm SN} v_{\rm t}^{-5} t^{-3}
	\left\{
	\begin{array}{ll}
	1 & {\rm for} ~ r < v_{\rm t} t \\
		\left( v_{\rm t} t / r \right)^{\omega} & {\rm for} ~ r > v_{\rm t} t
	\end{array}
	\right.,
\end{eqnarray}
where $v_{\rm t} = \sqrt{10 (\omega - 5) E_{\rm SN} / (3 (\omega - 3) M_{\rm ej})}$ and we assume $\omega = 9$.
As the shell expands, it eventually collides with the reverse shock of the SNR at a time $t = t_{\rm coll}$ ($\sim$ a few thousand years in this study), and we do not consider $t > t_{\rm coll}$ in the present paper.
It has already been pointed out by \citet{Bandiera+20} that the hydrodynamical properties of the reverse shocked SNR required for $t > t_{\rm coll}$ have no simple analytical expression.

The pressure inside the PWN $P_{\rm PWN}(t)$ is the sum of the contribution from the accelerated particles and the magnetic field.
The particle pressure inside the PWN is one-third of the total particle energy $\int N(\gamma, t) \gamma m_{\rm e} c^2 d \gamma$ divided by the volume of the PWN, where the energy distribution of the accelerated particles (electrons / positrons) $N(\gamma, t)$ evolves with time according to the stochastic acceleration model described in the next section \ref{sec:Acceleration}.
For the evolution of the magnetic field inside the PWN $B_{\rm PWN}(t)$, we simply assume \citep[e.g.,][]{Tanaka&Takahara10}
\begin{eqnarray}\label{eq:MagneticEvolution}
    \frac{4 \pi}{3} R_{\rm sh}^3(t) \frac{B_{\rm PWN}^2(t)}{8 \pi}
	& = &
	\eta_{\rm B} \int^t_0 L_{\rm spin}(t') d t', \nonumber \\
	& \equiv &
	\eta_{\rm B} E_{\rm rot}(t)
\end{eqnarray}
where $L_{\rm spin}(t) = L_0 (1 + t/\tau_0)^{-(n+1)/(n-1)}$ is the spin-down power of the pulsar powering the PWN, $n$ is the braking index, $L_0$ is the initial spin-down power, $\tau_0$ is the spin-down time and $E_{\rm rot}(t)$ is the injected rotational energy at a time $t$.
Note that $L_0 = 8 \pi^4 B_0^2 R^6_{\rm NS}/ (3 c^3 P^4_0)$ and $\tau_0 = 6 M_{\rm NS} c^3 P^2_0 / (20 \pi^2 B^2_0 R^4_{\rm NS})$ are also written in term of the initial period $P_0$, initial surface magnetic field $B_0$, radius $R_{\rm NS}$ and mass $M_{\rm NS}$ of the neutron star.
We introduced the magnetic fraction parameter $\eta_{\rm B}$, which is typically much smaller than unity for young PWNe \citep{Tanaka&Takahara11, Tanaka&Takahara13a}.
The rest of the spin-down power is divided into the accelerated particles injected directly into the PWN and the energy of the turbulence accelerating the radio-emitting particles.

\subsection{Stochastic Acceleration}\label{sec:Acceleration}

The turbulence, which is inevitably excited inside the PWN by the Rayleigh-Taylor instability at the contact surface between PWN and SNR \citep[e.g.,][]{Suzuki&Maeda19} and also by the kink instability \citep[e.g.,][]{Porth+14}, causes the stochastic acceleration.
The energy distribution of the accelerated particles $N(\gamma, t)$ is obtained by solving the Fokker-Planck equation
\begin{eqnarray}\label{eq:FokkerPlanck}
	\frac{\partial}{\partial t} N +\frac{\partial}{\partial \gamma} \left[\left(\dot{\gamma}_{\rm cool}(\gamma, t) - \gamma^2 D_{\gamma \gamma}(\gamma, t) \frac{\partial}{\partial \gamma} \frac{1}{\gamma^{2}}\right) N \right] \nonumber \\
	= Q_{\rm ext}(\gamma, t) + Q_{\rm PSR}(\gamma, t),
\end{eqnarray}
where $\gamma$ is the Lorentz factor of the electrons / positrons and $\dot{\gamma}_{\rm cool}$ is the cooling term including the adiabatic, synchrotron and the inverse Compton cooling.
The synchrotron radiation, the inverse Compton scattering off the cosmic microwave background radiation (IC/CMB), and off the synchrotron radiation (SSC) are calculated from $N(\gamma,t)$ \citep[c.f.,][]{Tanaka&Takahara10}.

The stochastic acceleration is described by the diffusion coefficient in the momentum space $D_{\gamma \gamma}$ in equation~(\ref{eq:FokkerPlanck}).
We assume the form of
\begin{eqnarray}\label{eq:DiffusionCoefficient}
    D_{\gamma \gamma}(\gamma, t) = \frac{\gamma^2}{2 \tau_{\rm acc}} \exp \left(-\frac{t}{\tau_{\rm turb}}\right), 
\end{eqnarray}
where $\tau_{\rm acc}$ is the initial acceleration time and $\tau_{\rm turb}$ is the decay time-scale of the turbulence \citep[c.f.,][]{Tanaka&Asano17}.
Equation~(\ref{eq:DiffusionCoefficient}) is the hardsphere formula considering the non-resonant interaction between turbulence and particles \citep[e.g.,][]{Ptuskin88}, and then the acceleration time-scale $t_{\rm acc}$ is independent from the particle Lorentz factor as
\begin{eqnarray}\label{eq:AccelerationTime}
	t_{\rm acc}(t) \equiv \frac{\gamma^2}{2 D_{\gamma \gamma}} = \tau_{\rm acc} \exp{\left(\frac{t}{\tau_{\rm turb}}\right)}. 
\end{eqnarray}
$t_{\rm acc}(t)$ would be related with the strength of the turbulence.
The turbulence decays when most of the turbulent energy is transferred to the accelerated particles by the stochastic acceleration, where the energy of the stochastically accelerated particles appears to be a non-negligible fraction of the rotational energy $E_{\rm rot}(t)$.
The time-scales $(\tau_{\rm acc}, \tau_{\rm turb})$ are the main parameters of the present model representing the time evolution of the turbulence inside the PWN.
Equation~(\ref{eq:DiffusionCoefficient}) is simpler than the diffusion coefficient introduced in \citet{Tanaka&Asano17} and has no (artificial) cut-off Lorentz factor for the stochastic acceleration (see also the discussion in section \ref{sec:Dis&Cons}).

Another important ingredient to reproduce the `flat radio spectrum' is the external sources of the radio-emitting particles $Q_{\rm ext}$ in equation~(\ref{eq:FokkerPlanck}), where the external sources are also expected from the independent study of the pair cascade process inside the pulsar magnetosphere \citep[e.g.,][]{Timokhin&Harding15}.
\citet{Tanaka&Asano17} considered that two possible origins of the external particle source.
One is that at the very early stage ($\sim$ weeks to months) of PWN evolution when the PWNe is thick from the pair creation process, a fraction of the pulsar's rotational energy is converted into the non-relativistic pairs, we call this the `impulsive injection' case.
The other is that the non-relativistic electrons are continuously supplied from the surrounding SN ejecta through the contact surface between the PWN and the SNR, which we call the `continuous injection' case.
We consider both of the impulsive and the continuous injection cases for $Q_{\rm ext}$ as
\begin{eqnarray}
	Q_{\rm ext}(\gamma, t) =
	\left\{
   	\begin{array}{ll}
	q_{\rm imp}(t) \\
	q_{\rm cont}(t) \\
	\end{array} \right\}
	\delta(\gamma - \gamma_{\rm inj}),
\end{eqnarray}
where the injection energy is $\gamma_{\rm inj} \sim 1$ \citep[][]{Tanaka&Asano17}.
The observed radio flux constrains the values of $q_{\rm imp}(t)$ or $q_{\rm cont}(t)$ because the radio synchrotron flux is proportional to the amount of the externally injected particles.

For the impulsive injection $q_{\rm imp}$, the low energy particles are present from the beginning of the calculations of $t_{\rm init} =$ 1 yr.
A small fraction $f_{\rm imp} \ll 1$ of the injected rotational energy at $t = t_{\rm init}$ ($E_{\rm rot}(t_{\rm init}$)) would be converted into the electron-positron pair inside the newborn PWN surrounded by the thick SNR, we set
\begin{eqnarray}\label{eq:ImpulsiveInjection}
	q_{\rm imp}(t) = f_{\rm imp} \frac{E_{\rm rot}(t_{\rm init})}{m_{\rm e} c^2} \delta(t - t_{\rm init}).
\end{eqnarray}
Instead of $f_{\rm imp}$, the time-scale $\tau_{\rm pair}$ which satisfies $E_{\rm rot}(\tau_{\rm pair}) \equiv f_{\rm imp} E_{\rm rot}(t_{\rm init})$ or the number of pairs $N_{\rm pair} \equiv f_{\rm imp} E_{\rm rot}(t_{\rm init}) / m_{\rm e} c^2$ can be a measure of the pair amount (see also the discussion in section \ref{sec:Dis&Cons}).

For the continuous injection from the SNR $q_{\rm cont}$, we set $q_{\rm cont}(t) \propto \dot{M}_{\rm sh}(t)$.
The neutral components of the SN ejecta can easily penetrate into the PWN and be photoionized subsequently, for example.
Considering that a small fraction of the swept-up mass $f_{\rm cont} \ll 1$, e.g., the particles of the suprathermal tail of the momentum distribution, will be injected into the stochastic acceleration process, i.e.,
\begin{eqnarray}
	q_{\rm cont}(t) = f_{\rm cont} \frac{\dot{M}_{\rm sh}(t)}{m_{\rm p}}
\end{eqnarray}
where $m_{\rm p}$ is the mass of a proton.
Although the same number of hadrons can be injected for the continuous injection case, we assume that the stochastically accelerated hadrons have at most the same order of energy as the electrons, i.e. the system could have an energy a factor of two larger than the present study.
We will discuss the total energy inside PWN 1986J in section \ref{sec:Dis&Cons}.
We ignore the acceleration and emission processes of the hadrons in the present paper.

Finally, for the particle injection from the pulsar $Q_{\rm PSR}$, we assume the single power-law distribution characterized by the minimum and maximum Lorentz factors $\gamma_{\rm min}, \gamma_{\rm max}$, and the power-law index $p$, i.e.,
\begin{eqnarray}\label{eq:InjectionFromPSR}
	Q_{\rm PSR}(\gamma, t) = \dot{n}_{\rm PSR}(t) \gamma^{-p} ~ {\rm for} ~ \gamma_{\rm min} < \gamma < \gamma_{\rm max}.
\end{eqnarray}
The normalization $\dot{n}_{\rm PSR}$ is determined to satisfy $\int d \gamma Q_{\rm PSR}(\gamma, t) \gamma m_{\rm e} c^2 d \gamma=\eta_{\rm e} L_{\rm spin}(t)$, where $\eta_{\rm e}$ is the energy fraction of the pulsar wind particles accelerated at the termination shock of the wind.
For $\gamma_{\rm max}$, the synchrotron cooling limit $\gamma_{\rm max, syn} \propto B_{\rm PWN}^{-1/2}$ is imposed at each time step \citep[e.g.,][]{deJager+96}, i.e., the maximum energy of the synchrotron photon is limited by $\approx m_{\rm e} c^2 / \alpha_{\rm f}$, where $\alpha_{\rm f}$ is the fine structure constant.

Note that all injected particles $N(\gamma, t)$, both $Q_{\rm ext}(\gamma, t)$ and $Q_{\rm PSR}(\gamma, t)$, are subject to the stochastic acceleration (and the coolings) according to equation (\ref{eq:FokkerPlanck}).
The acceleration time-scale is the same for all the particles, while the radiative cooling is more effective for the higher particle energies.
As a result, only the (low-energy) externally injected particles $Q_{\rm ext}$ are predominantly accelerated to become the radio-emitting particles, i.e., the particles of $\gamma \lesssim 10^5$ in the lower-right panels of Figs. \ref{fig:Impulsive} and \ref{fig:Continuous}.
In the present model, the acceleration time-scale $t_{\rm acc}(t)$ increases exponentially for $t > \tau_{\rm turb}$ (see equation (\ref{eq:AccelerationTime})), i.e., the particle acceleration continues until the age of the system becomes $\sim \tau_{\rm turb}$ and must be stopped by the decay of the turbulence.
The decay mechanism of the turbulence is not modelled here.
Since equations \ref{eq:FokkerPlanck} and \ref{eq:DiffusionCoefficient} do not guarantee the energy conservation for any set of $\tau_{\rm acc}$ and $\tau_{\rm turb}$, they need to be chosen within a range that does not violate energy conservation (see, however, the discussion in section \ref{sec:Dis&Cons}).

\subsection{Fitting Procedure}\label{sec:FittingProcedure}

The main parameters to be studied in the present paper are those of the diffusion coefficient $D_{\gamma \gamma}$ $(\tau_{\rm acc}, \tau_{\rm turb})$ and of the external particle injection $Q_{\rm ext}$ $(f_{\rm imp}, f_{\rm cont})$.
In addition, the system also has the parameters of the SN ejecta and the surrounding interstellar medium $(E_{\rm SN}, M_{\rm ej}, n_{\rm ISM})$, of the central pulsar $(P_0, B_0, M_{\rm NS}, R_{\rm NS}, n)$, of the particle spectrum $Q_{\rm PSR}$ $(p, \gamma_{\rm min}, \gamma_{\rm max}, \eta_{\rm e})$, and of the magnetic energy fraction of the spin-down power $\eta_{\rm B}$.
In the present study, we fix the latter parameters and focus on whether or not the well-known Crab-like system can reproduce the radio light curve of PWN 1986J.
For this purpose, we adopt most of the parameters from the Crab Nebula and we fit only the main parameters $(\tau_{\rm acc}, \tau_{\rm turb}, f_{\rm imp}, f_{\rm cont})$ to reproduce both the radio light curve of PWN 1986J ($\sim$ 30 yr) and the broadband spectrum to be the Crab Nebula ($\sim$ kyr).
The parameters of the remnant of SN 1986J are $E_{\rm SN} = 10^{51}~{\rm erg},\ M_{\rm ej} = 9.5~M_{\odot},$ and $n_{\rm ISM} = 0.1~{\rm cm^{-3}}$ \citep[c.f.][]{MacAlpine&Satterfield08, Bucciantini+11}.
The parameters of the spin-down evolution of PSR 1986J are $P_0 = 18.7$ msec, $B_0 = 3.35 \times 10^{12}$ G, $n = 2.51$, $M_{\rm NS} = 1.4~M_{\odot}$ and $R_{\rm NS} = 12$ km ($L_0 = 2.6 \times 10^{39}\ {\rm erg ~ s^{-1}},~\tau_0 = 1.1$ kyr).
The parameters of $B_{\rm PWN}(t)$ and $Q_{\rm PSR}(\gamma,t)$ are also almost the same as \citet{Tanaka&Asano17}, i.e., $\eta_{\rm B} = 5 \times 10^{-3}, \eta_{\rm e} = 0.5, p = 2.5, \gamma_{\rm min} = 4 \times 10^5$, and $\gamma_{\rm max} = 7 \times 10^9$.
The adopted parameters are summarized in Table~\ref{tbl:Parameters}.
Note that $n_{\rm ISM}$ is relatively small for SN IIn but this does not affect the behaviors in the early phase of the evolution ($\sim$ 30 yr).

The time-scales of the diffusion coefficient ($\tau_{\rm acc}, \tau_{\rm turb}$) and the injection efficiency to the stochastic acceleration process ($f_{\rm imp}$ or $f_{\rm cont}$) are only parameters which are loosely determined by \citet{Tanaka&Asano17}.
The fraction parameters $f_{\rm imp}$ or $f_{\rm cont}$ are rather well determined because they are chosen to reproduce the observed radio flux of the Crab Nebula and to account for the total amount of particles in the Crab Nebula \citep[][]{Tanaka&Asano17}.
They showed that $\tau_{\rm acc}$ of a few tens and $\tau_{\rm turb}$ of a few hundreds of years reproduce the observed flat radio spectrum of the Crab Nebula, but the other sets of $(\tau_{\rm acc}, \tau_{\rm turb})$ would be allowed (see also the discussion in section \ref{sec:Dis&Cons}).
We are going to find a set of $(\tau_{\rm acc}, \tau_{\rm turb})$ that reproduce the observed properties of both PWN 1986J (5 GHz light curve) and the Crab Nebula (broadband spectrum at 1 kyr).
Roughly speaking, $\tau_{\rm acc}$ and $\tau_{\rm turb}$ determine the rise and fall times of the radio light curve, respectively.
For example, the calculated radio light curve will rise/fall faster for the shorter acceleration ($\tau_{\rm acc}$)/decay ($\tau_{\rm turb}$) time-scales.
One-year and five-year resolutions are used to find the best-fit values of $\tau_{\rm acc}$ and $\tau_{\rm turb}$, respectively.

\section{Application}\label{sec:Application}

%
\begin{table}
\caption{
	Summary of the parameters for the calculations in Figs~\ref{fig:Impulsive} and \ref{fig:Continuous}, corresponding to the cases of the impulsive and continuous injection of the radio-emitting particles, respectively.
}
\label{tbl:Parameters}
\begin{center}
\begin{tabular}{cccc}
	\multicolumn{2}{c}{Symbol}  &
	Impulsive &
	Continuous  \\
\hline
\multicolumn{4}{c}{Fitted Parameters}    \\
\hline
	\multicolumn{2}{c}{$\tau_{\rm acc}$ (yr)}  & 10 & 9  \\
	\multicolumn{2}{c}{$\tau_{\rm turb}$ (yr)} & 75 & 70  \\
	\multicolumn{2}{c}{$f_{\rm imp}$}          & $2.0 \times 10^{-2}$ & $\cdots$   \\
	\multicolumn{2}{c}{$f_{\rm cont}$}         & $\cdots$ & $6.0 \times 10^{-5}$   \\
\hline
\multicolumn{4}{c}{Parameters Adopted from The Crab} \\
\hline
	$E_{\rm SN}$       & $10^{51}$ erg  &	$M_{\rm ej}$       & 9.5 $M_{\odot}$  \\
	$n_{\rm ISM}$      & 0.1 cm$^{-3}$  & $d$                & 10 Mpc   \\
	$M_{\rm NS}$       & 1.4 $M_{\odot}$  &	$R_{\rm NS}$       & 12 km   \\
	$P_0$              & 18.7 msec &	$B_0$              & $3.35 \times 10^{12}$ G \\
	$n$                & 2.51 &	$p$                & 2.5  \\
%
%
	$\eta_{\rm B}$     & 5.0 $\times 10^{-3}$ &	$\eta_{\rm e}$     & 0.5  \\
	$\gamma_{\rm min}$ & $4.0 \times 10^{5}$  &	$\gamma_{\rm max}$ & 7.0 $\times 10^{9}$  \\
\end{tabular}
\end{center}
\end{table}
%
%
\begin{figure*}
\begin{center}
	\includegraphics[scale=0.6]{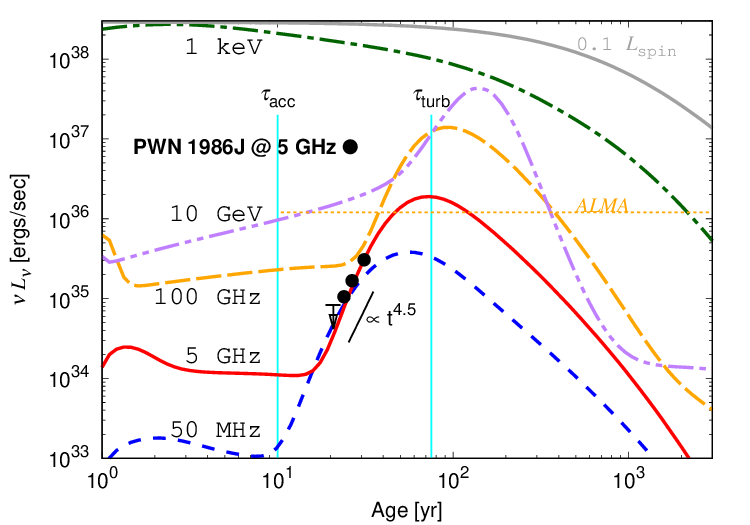}
	\includegraphics[scale=0.6]{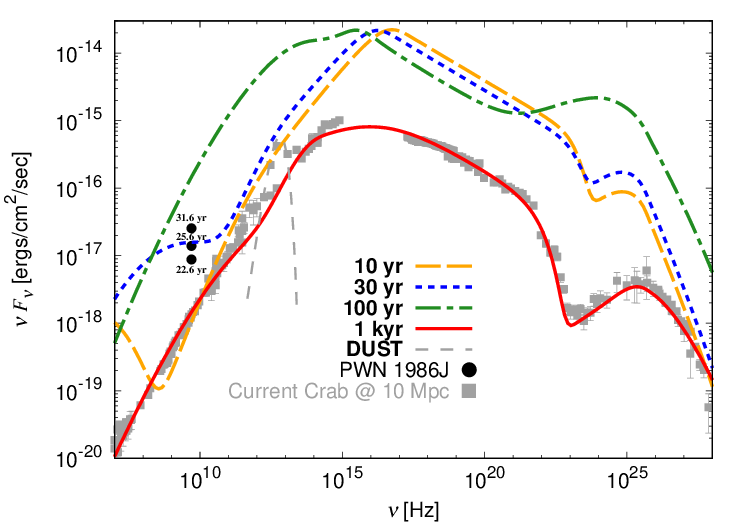}
	\includegraphics[scale=0.6]{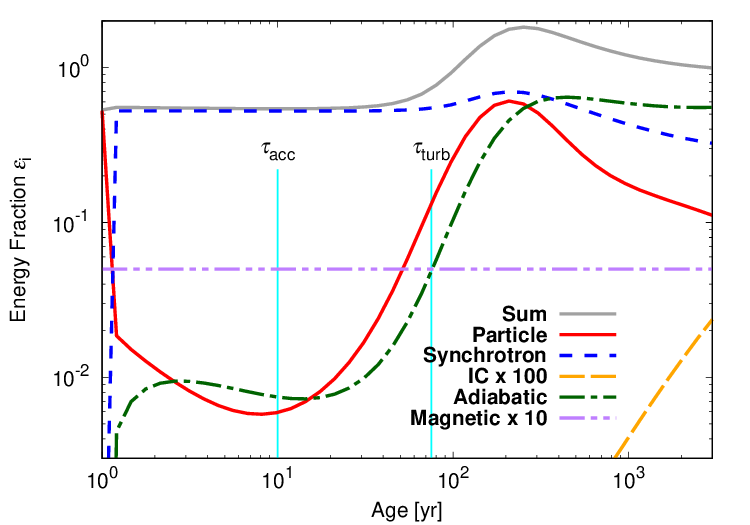}
	\includegraphics[scale=0.6]{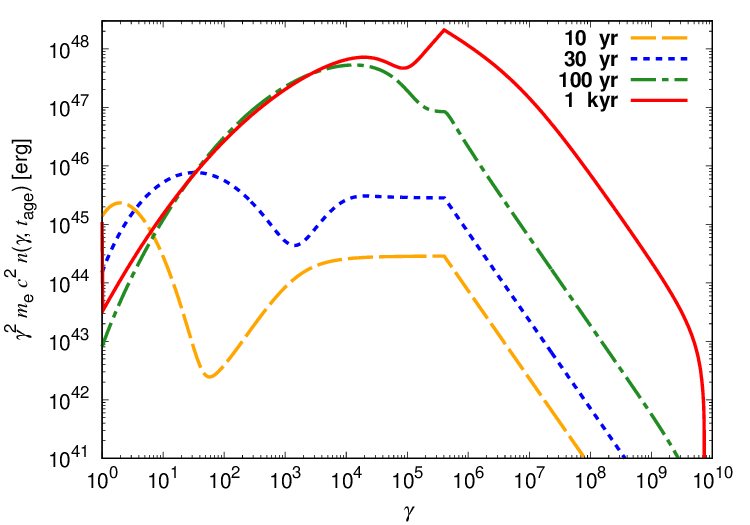}
\end{center}
\caption{
	Results for the impulsive injection case.
	The adopted parameters are summarized in Table~\ref{tbl:Parameters}.
	Top-left: the $\le$ 3 kyr light curve at 50 MHz (blue dotted), 5 GHz (red solid), 100 GHz (yellow dashed), 1 keV (green dot-dashed), and 10 GeV (purple dot-dot dashed) with the observational data of PWN 1986J assuming 10 Mpc \citep{Bietenholz&Bartel17a}.
	The grey solid line shows one tenth of the spin-down power $0.1 L_{\rm spin}(t)$ and the vertical cyan lines correspond to the adopted values of $\tau_{\rm acc}$ and $\tau_{\rm turb}$, respectively.
	The horizontal yellow dotted line is the sensitivity line for {\it ALMA} (Band 3) $\sim 0.1$ mJy taken from \citet{Murase+21}.
	Top-right: the broadband photon spectra at $t_{\rm age} =$ 10 yr (yellow dashed), 30 yr (blue dotted), 100 yr (green dot-dashed) and 1 kyr (red solid).
	The 5 GHz data points at 22.6, 25.6 and 31.6 yr are shown in black circles.
	The grey observational data points of the Crab Nebula (placed 10 Mpc away from us) are taken from \citet{Baldwin71, Baars+77, Macias-Perez+10} for radio, \citet{Ney&Stein68, Grasdalen79, Green+04, Temim+06} for IR and optical, \citet{Kuiper+01} for X-rays, and \citet{Aharonian+06, Albert+08, Abdo+10} for $\gamma$-rays.
	Bottom-left: time evolution ($\le$ 3 kyr) of the fractional energy of the particles $\int d \gamma \gamma m_{\rm e} c^2 N(\gamma, t)$ (red solid), of the radiated and wasted ones by synchrotron (blue dotted), inverse Compton off the CMB (yellow dashed, $10^2$ times amplified) and adiabatic (green dot-dashed) cooling $\int^t_0 d t'\int d \gamma |\dot{\gamma}_i(\gamma, t')| m_{\rm e} c^2 N(\gamma, t')$ ($i =$ Synch, IC/CMB, Adiabatic), of the magnetic field $\eta_{\rm B} E_{\rm rot}(t)$ (purple dot-dot dashed, ten times amplified) and of the sum of all of them (thin black).
	Bottom-right: the particle energy spectra at $t_{\rm age} =$ 10 yr (yellow dashed), 30 yr (blue dotted), 100 yr (green dot-dashed) and 1 kyr (red solid).
}
\label{fig:Impulsive}
\end{figure*}
%

%
\begin{figure*}
\begin{center}
	\includegraphics[scale=0.6]{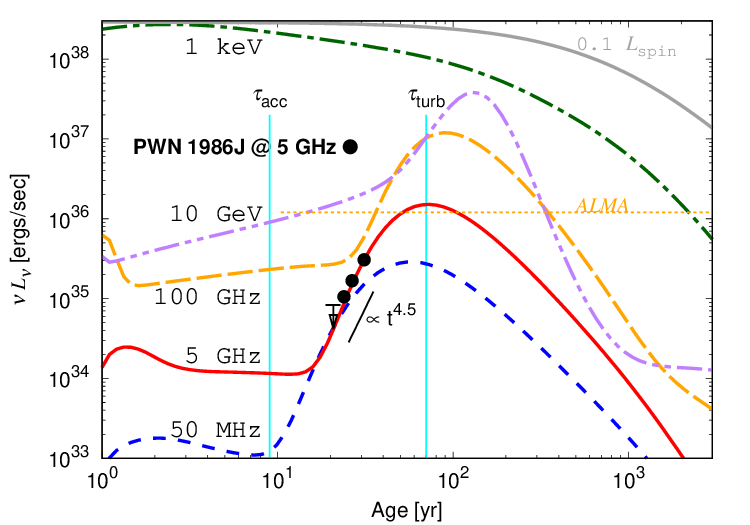}
	\includegraphics[scale=0.6]{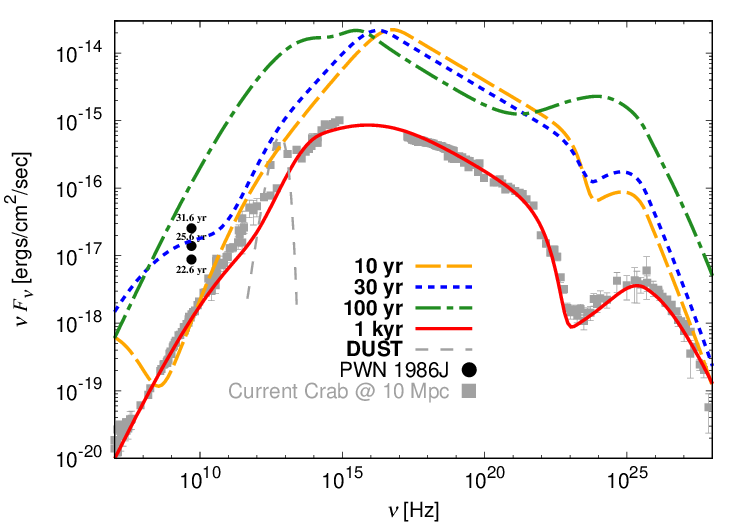}
	\includegraphics[scale=0.6]{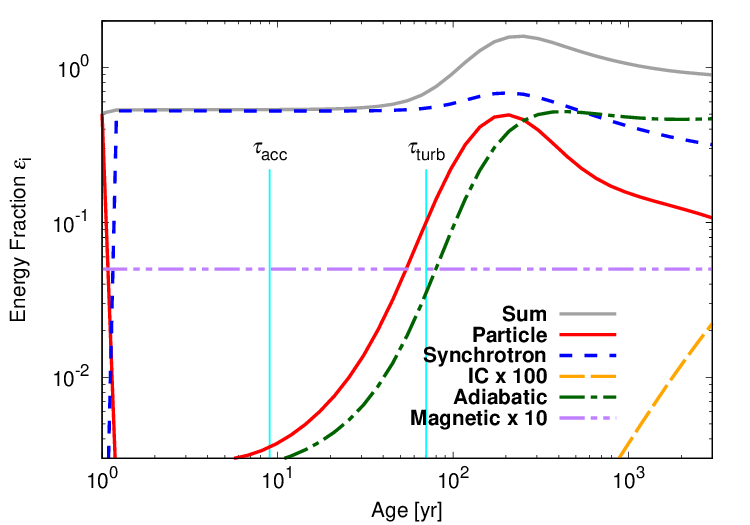}
	\includegraphics[scale=0.6]{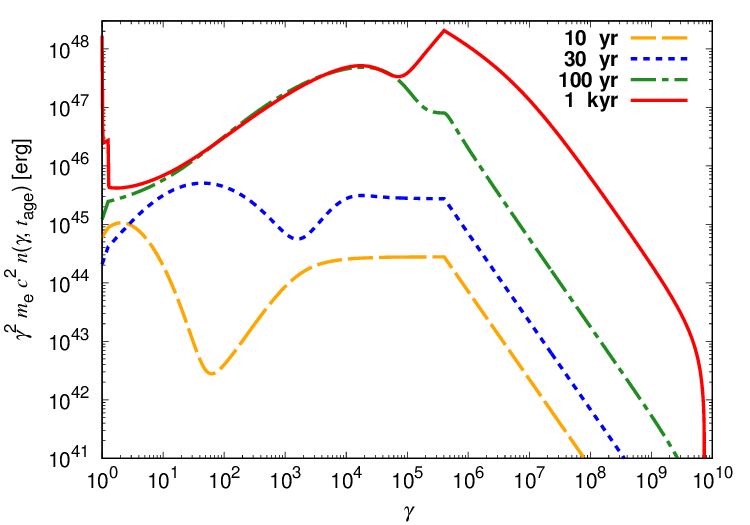}
\end{center}
\caption{
	Results for the continuous injection case, where the multi-band light curves (top-left), the broadband photon spectra (top-right), evolution of the fractional energy (bottom-left) and the particle spectra (bottom-right).
	See the caption in Fig.~\ref{fig:Impulsive} for details.
	The used parameters are summarized in Table~\ref{tbl:Parameters}.
}
\label{fig:Continuous}
\end{figure*}

We assume that the central component of SN 1986J is a PWN powered by the relic neutron star SN 1986J.
We adopt the distance of 10 Mpc to SN 1986J (NGC 891) \citep[e.g.,][]{Tonry+01}.
The central component shows an elliptical shape with the major axis FWHM $\approx 0.022$ pc at an age of $\approx$ 30 yr and is consistent with no proper motion \citep[][]{Bietenholz&Bartel17a}.
The continuous VLBI observations of SN 1986J give the multi-frequency radio light curve \citep[][]{Bietenholz+04, Bietenholz+05, Bietenholz&Bartel17b}.
\citet{Bietenholz&Bartel17a} gave the 5 GHz light curves of both the central (PWN) and shell (SNR) components separately and showed that the 5 GHz flux keeps increasing only for the central component (PWN 1986J).
Below, we show the results for the impulsive (Fig.~\ref{fig:Impulsive}) and continuous (Fig.~\ref{fig:Continuous}) injection cases, separately.


The top-left panel of Fig.~\ref{fig:Impulsive} (impulsive injection) shows that the multi-band light curve.
The observed 5 GHz light curve of PWN 1986J (red solid) is reproduced with $\tau_{\rm acc} = 10~{\rm yr},\ \tau_{\rm turb} = 75~{\rm yr}$ and the injection efficiency to the stochastic acceleration process $f_{\rm imp} = 2 \times 10^{-2}$ (equivalent to $\tau_{\rm pair} \approx$ one week and also $N_{\rm pair} \approx 2 \times 10^{51}$).
The radio (50 MHz, 5 GHz, 100 GHz) light curves show similar rising ($10 \lesssim t \lesssim 20$ yr) and decaying ($50 \lesssim t \lesssim 100$ yr) behavior corresponding to $\tau_{\rm acc} = 10$ yr and $\tau_{\rm turb} = 75$ yr because the stochastic acceleration is responsible for the radio-emitting particles.
The X-ray (1 keV) light curve has similar evolution to the spin-down power (grey solid) because the X-ray-emitting particles are supplied by PSR 1986J ($Q_{\rm PSR}$).
The $\gamma$-ray (10 GeV) light curve for $\lesssim 1$ kyr is in the SSC dominated regime, while the $\gamma$-ray luminosity is almost constant for $\gtrsim 1$ kyr in the IC/CMB dominated regime.

The broadband spectra of PWN 1986J at 10, 30, 100 and 1 kyr are shown in the top-right panel of Fig.~\ref{fig:Impulsive}.
PWN 1986J is about 30 yr (blue dotted) at present and the 5 GHz flux is reproduced (black stars).
We have chosen the parameters for the broadband spectral evolution of PWN 1986J to be consistent with that of the Crab Nebula at an age of 1 kyr (grey squares).
According to the rapid increase of the radio-emitting particles ($\gamma \lesssim 10^5$) due to the stochastic acceleration (yellow dashed, blue dotted and green dashed lines in bottom-right panel), the radio flux increases with time despite the rapid decrease of the magnetic field inside the nebula according to its expansion.
This is a characteristic feature of the stochastic acceleration model, while for conventional broken power-law injection models \citep[e.g.,][]{Tanaka&Takahara10, Bucciantini+11, Torres+14, Zhu+18}, the radio flux always decreases with time as shown by the 1 keV light curve in the top-left panel of Fig.~\ref{fig:Impulsive}.

The bottom-left panel of Fig.~\ref{fig:Impulsive} shows the evolution of the particle, magnetic, and radiated energies normalized by the total rotational energy injected by PSR 1986J at each time $E_{\rm rot}(t)$.
Until $\sim$ 100 yr, most of the injected rotational energy is radiated away from PWN 1986J by synchrotron radiation (blue dotted) and then the particle energy (red solid) and the adiabatic cooling (green dot-dashed) start to dominate.
The sum of each energy (thin black solid) slightly exceeds the total energy supply from the central pulsar $E_{\rm rot}(t)$ after $\sim$ 100 yr until a few kyr for our best-fit case of the 5 GHz light curve and the broadband spectrum.
This situation can be relaxed by introducing a slightly short initial period (e.g., $P_0 = 10$ msec) or by considering the supply of the turbulent energy from the surrounding SNR ejecta, but changing the pulsar parameters is beyond the scope of the present study (see also section \ref{sec:Dis&Cons}).
Here, we conclude that the slightly short period Crab-like pulsar can explain the 5 GHz light curve of PWN 1986J.

Fig.~\ref{fig:Continuous} shows the results for the case of the continuous particle injection from the SN ejecta, where we take $\tau_{\rm acc} = 9~{\rm yr},\ \tau_{\rm turb} = 70~{\rm yr}$ and the injection efficiency to the stochastic acceleration process $f_{\rm cont} = 6 \times 10^{-5}$.
The values of ($\tau_{\rm acc}, \tau_{\rm turb}$) are close to those of the impulsive injection case.
The multi-band light curves are also similar to each other (top-left panel) and then the injection models are difficult to distinguish from the spectral information alone.
In the particle distribution (bottom-right panel), the number of the low energy $\gamma < 10^2$ particles which do not contribute to the radiation spectrum above 10 MHz is different between the impulsive and continuous injection cases.
The time evolution of the energetics (bottom-left panel) is also similar to that of Fig.~\ref{fig:Impulsive}.

Finally, the expansion of the PWN $R_{\rm sh}(t)$ is similar to the results of \citet{Gelfand+09}, where the expansion velocity $\dot{R}_{\rm sh}$ is almost constant around 1000 km s$^{-1}$ for a few hundreds years and then accelerates to $\sim$ 2000 km s$^{-1}$ at $\sim$ 1 kyr.
The diameter of the calculated PWN at an age of 30 yr is $\approx 0.041$ pc which is about twice as large as the observed FWHM of PWN 1986J but still the same order.
Adjusting the size of the PWN by changing the parameters of the parent SN is also beyond the scope of the present study.
Still, we conclude that the Crab-like system can explain both the 5 GHz light curve and size of the radio counterpart of PWN 1986J.
Note that, for more precise comparison of the size of the nebula with the observed FWHM, we need to model the brightness profile of PWN 1986J.
The magnetic field strength at an age of 30 yr is $\approx$ 17 mG.
These values are almost the same for both the impulsive and the continuous injection cases.

\section{Discussion \& Conclusions}\label{sec:Dis&Cons}

The observed 5 GHz light curve of PWN 1986J (the central component of SN 1986J) is reproduced by the stochastic acceleration model of PWNe developed by \citet{Tanaka&Asano17}.
By setting the parameters of PSR 1986J (the central pulsar powering PWN 1986J) and of SN 1986J similar to those of the Crab Nebula, we found the sets of the stochastic acceleration time-scale and the decay time-scale of the turbulence in order to reproduce the 5 GHz light curve of PWN 1986J in addition to the broadband spectrum of the Crab Nebula at an age of one thousand years.
SN 1986J is a plausible candidate containing the youngest pulsar ever observed ($\sim$ 30 years) with the similar pulsar parameters as the Crab pulsar.

PWN 1986J is about twenty times more luminous than the current Crab Nebula at 5 GHz and this is simply explained by the strong magnetic field strength for a young (small) PWN, while the increasing radio flux is a unique feature of the stochastic acceleration model.
We require the stochastic acceleration time-scale of $\tau_{\rm acc} \approx 10$ yr to explain the observation of PWN 1986J.
The small value of $\tau_{\rm acc}$ compared to the age of the observed PWNe $\gtrsim$ kyr suggests that the strong turbulence inside the PWN was driven in a very early phase and had already decayed after about one hundred years ($\tau_{\rm turb} \approx$ 70 yr).
In other words, the stochastic acceleration phase can only be studied for very young PWNe, and PWN 1986J is the first candidate.
Further observations by {\it ALMA} and other mm/submm telescopes are very interesting not only to test the present model but also to study the spectral and dynamical role of the turbulence in PWNe \citep[c.f.,][]{Tanaka&Asano17, Tanaka+18}.

The fitted values of $f_{\rm imp}$ and $f_{\rm cont}$ in order to reproduce the 5 GHz flux of PWN 1986J also reproduce the current spectrum of the Crab Nebula, i.e., the amount of the radio-emitting particles is almost the same as \citet{Tanaka&Asano17}.
$f_{\rm imp} = 2 \times 10^{-2}$ implies $\tau_{\rm pair} \approx$ a week and $f_{\rm cont} \ll 1$ implies that only a small fraction of particles penetrating into the PWN is injected the stochastic acceleration process, although we do not repeat the further discussion about the external injection \citep[see section 4.2 of][]{Tanaka&Asano17}.
The present study also does not allowed to distinguish the impulsive (Fig.~\ref{fig:Impulsive}) and continuous (Fig.~\ref{fig:Continuous}) injection cases and then the details of the pair creation process in the very early phase of the evolution (impulsive injection) or the particle penetration process (continuous injection) should be studied.

We have only the 5 GHz light curve of PWN 1986J \citep[][]{Bietenholz&Bartel17a}.
\citet{Bietenholz&Bartel17b} interpreted the observed radio spectral evolution as the sum of the optically thin (SNR) component $S_{\rm SNR}$ and the optically thick (PWN) component $S_{\rm PWN, abs}$.
They found that the two component model of $S_{\rm SNR} \propto t^{-3.92} \nu^{-0.63}$ and $S_{\rm PWN} \propto t^{-2.07} \nu^{-0.76}$ with the optical depth for the free-free absorption (FFA) $\tau_{\rm FFA} \propto \nu^{-2.1} t^{-2.72}$ ($S_{\rm PWN, abs} = S_{\rm PWN} (1 - e^{-\tau_{\rm FFA}}) / \tau_{\rm FFA}$) can reproduce the evolution of the multi-band radio spectrum.
However, it is easy to find that $S_{\rm PWN,abs}(5~{\rm GHz},t) \sim S_{\rm PWN} / \tau_{\rm FFA} \propto t^{0.65}$ does not fit to the observed 5 GHz light curve $\propto t^{4.5}$ (see top-right panel of Fig.~\ref{fig:Impulsive}), where $\tau_{\rm FFA}(5~{\rm GHz},t) \gg 1$ for $t \le 30$ yr in their model.
The spectral decomposition of the PWN component from the SNR one \citep[][]{Bietenholz&Bartel17b} is not consistent with the spatial decomposition \citep[][]{Bietenholz&Bartel17a}.
Spatially decomposed multi-band light curves help to understand the nature of PWN 1986J.

The assumption of an optically thin 5 GHz emission from PWN 1986J in the present model is justified as follows.
For FFA through the surrounding ejecta, the optical depth is \citep[e.g.,][]{Lang99}
\begin{eqnarray}\label{eq:FFA}
	\tau_{\rm FFA}
	& \approx &
	1.12 \times 10^{-8}
	\left( \frac{\nu}{5~{\rm GHz}} \right)^{-2.1} \nonumber \\
	& \times &
	\left( \frac{T_{\rm e}}{10^4~{\rm K}} \right)^{-1.35}
	\left( \frac{{\rm EM}}{{\rm cm^{-6} pc}} \right).
\end{eqnarray}
where EM is the emission measure.
For example, we obtain $R_{\rm sh} \approx 0.013$ pc and $R_{\rm SNR} \approx 0.36$ pc at an age of 20 yr from the present calculation (about a factor of two different from the SN 1986J system).
Considering the extreme case that the ejecta is fully ionized protons,
\begin{eqnarray}\label{eq:EjectaNumberDensity}
	n_{\rm ej}
	& = &
	\frac{M_{\rm ej}}{m_{\rm p}} \left[ \frac{4 \pi}{3} (R_{\rm SNR}^3 - R_{\rm sh}^3) \right]^{-1} \nonumber \\
	& \approx & 
	1.6 \times 10^{3}~{\rm cm^{-3}},
\end{eqnarray}
the optical depth for FFA is $\tau_{\rm FFA} \approx 10^{-2}$.

For the synchrotron self-absorption (SSA) inside PWN 1986J, the brightness temperature of PWN 1986J can be estimated from the relation \citep[e.g.,][]{Rybicki&Lightman79}
\begin{eqnarray}\label{eq:BrightnessTemperature}
	k_{\rm B} T_{\rm b}
	=
	\frac{c^2 d^2}{2 \pi \nu^2 R^2} F_{\nu}
	& \approx &
	15~{\rm keV}
    \left( \frac{\nu}{5~{\rm GHz}} \right)^{-2}
	\left( \frac{F_{\nu}}{511~{\rm \mu Jy}} \right) \nonumber \\
 	& \times &
	\left( \frac{R}{0.011~{\rm  pc}} \right)^{-2}
 	\left( \frac{d}{10~{\rm Mpc}} \right)^{2},
\end{eqnarray}
where we adopted the distance to the object $d = 10$ Mpc, the radius of the spherical nebula $0.011$ pc (a half of the FWHM of PWN 1986J), and the 5 GHz flux density 511 $\mu$Jy at an age of 31.6 yr from \citet{Bietenholz&Bartel17b}.
The value of the brightness temperature is much lower than the energy of a particle emitting 5 GHz synchrotron radiation and then the 5 GHz emission is not from an SSA-thick object.
For the higher frequency emission, we have to consider the bound-free/bound-bound absorption processes.
Although we do not discuss them in detail here, the absorption processes are inefficient at an age of at least 30 yr \citep[c.f.,][]{Metzger+14, Murase+15}.

The present model considers the radio emission from the optically thin region and is characterized by the slower peaks of the light curves for the higher frequency (top-left panels of Figs.~\ref{fig:Impulsive} and \ref{fig:Continuous}).
On the other hand, the optically thick model is required to explain the rising light curve of PWN 1986J within the conventional broken power-law model, where the radio light curve decreases with time like the X-ray one and is characterized by the slower peak at lower frequency because $\tau_{\rm FFA}$ is the decreasing function of a frequency.
The submillimeter observations of PWN 1986J with {\it ALMA} can distinguish between the optically thin and thick models.
Note that, although the X-ray emission from PWN 1986J seems detectable level with the current X-ray telescopes, the remnant of SN 1986J is much brighter than the PWN 1986J in X-ray \citep[][]{Houck+98}.
To clarify the detectability of PWN 1986J in X-ray, an X-ray radiative transfer calculation through the remnant is required, which is beyond the scope of the present paper.

The total energy inside PWN 1986J is slightly larger than $E_{\rm rot}(t)$ for $t >$ a few hundred years (bottom-left panels in Figs.~\ref{fig:Impulsive} and \ref{fig:Continuous}).
In addition, for the continuous injection case, we should also include the energy of the accelerated hadrons, which is of the same order of magnitude as that of the accelerated electrons.
This situation can be relaxed by introducing a slightly shorter $P_0$ and/or a slightly larger $B_0$ than the present values, as mentioned in section \ref{sec:Application}.
Note that we still do not need an extreme pulsar like $P_0 \le 10$ msec or $B_0 \ge 10^{13}$ G to achieve the consistency of the energetics, and the pulsar associated with PWN 1986J would be a Crab-like pulsar.
A more sophisticated model of the stochastic acceleration taking into account the evolution of the turbulence strength should be developed in future studies to further discuss the energetics of the stochastic acceleration model \citep[e.g.,][]{Kakuwa16}.

Finally, we adopted $t_{\rm init} = 1$ yr throughout the present paper.
Calculations starting from a smaller $t_{\rm init}$ are practically difficult for our explicit solver of equation~(\ref{eq:FokkerPlanck}).
However, the results obtained would not so sensitive to the choice of $t_{\rm init}$ unless $t_{\rm init} < \tau_{\rm acc}$ because the energy injected into the PWN is immediately radiated away by the synchrotron cooling, as can be seen in the bottom-right panels of Figs.~\ref{fig:Impulsive} and \ref{fig:Continuous}.
The energy of the synchrotron radiation would be acquired by the SN ejecta in a very early phase of evolution but the amount of $E_{\rm rot}$ is much smaller than $E_{\rm SN}$ in this `non-extreme pulsar' case.

\section*{Acknowledgements}

S. J. T. would like to thank S. Kisaka, K. Murase, T. Tezuka, S. Kunugi, A. Mikumo and K. Mita for useful discussion.
This work is supported by Grants-in-Aid for Scientific Research No. 20K04010(KK), 20H01904(KK), 22H00130(KK) and by JPJS Bilateral Program, Grant No. JPJSBP120229940.
S. J. T. would like to thank Aoyama Gakuin University Research Institute grant program for creation of innovative research, the Sumitomo Foundation, and the Research Foundation For Opto-Science and Technology for support.

\section*{Data availability}

The data underlying this article will be shared on reasonable request to the corresponding author.



\bibliographystyle{mnras}
\bibliography{draft} 


%
%
%
%
%

\bsp	
\label{lastpage}
\end{document}